\documentclass{IEEEtran}
\usepackage{cite}
\usepackage{amsmath,amssymb,amsfonts}
\usepackage{algorithmic}
\usepackage{graphicx}
\usepackage{textcomp}
\usepackage{xcolor}
\usepackage{enumitem}
\usepackage[numbers]{natbib}
\usepackage{booktabs}
\usepackage{float}

\newcommand{\Real}{\mathbb{R}}
\newcommand{\xh}{\hat{x}}
\newcommand{\yh}{\hat{y}}
\newcommand{\xt}{\tilde{x}}

\newcommand{\Ban}{B_\perp}

\newcommand{\Ean}{E_\perp}
\newcommand{\Xm}{\mathbf{X}} 

\newcommand{\dt}{\partial t}
\newcommand{\fb}{f_\partial}
\newcommand{\eb}{e_\partial}
\newcommand{\ze}{\zeta}

\newcommand{\Lo}{\mathcal{H}}
\newcommand{\Bo}{\mathcal{B}}
\newcommand{\Co}{\mathcal{C}}

\newcommand{\qeq}{q^\ast}
\newcommand{\peq}{p^\ast}
\newcommand{\Qeq}{Q^\ast}
\newcommand{\ueq}{u^\ast}
\newcommand{\yeq}{y^\ast}

\newtheorem{theo}{Theorem}
\newtheorem{proposition}{Proposition}
\newtheorem{remark}{Remark}

\newtheorem{corollary}[theo]{Corollary}
\newenvironment{Proof}[1][Proof]{\textbf{#1.} }{\ \rule{0.5em}{0.5em}}

\definecolor{jesus}{rgb}{0, 0, 1}

\definecolor{R1C-1a}{rgb}{0, 0, 0}

\definecolor{R1C-2}{rgb}{0, 0, 0}

\definecolor{R2C-3}{rgb}{0, 0, 0}

\newcommand{\violet}[1]{\textcolor{black}{#1}}
\newcommand{\jesus}[1]{\violet{#1}}

\def\BibTeX{{\rm B\kern-.05em{\sc i\kern-.025em b}\kern-.08em
    T\kern-.1667em\lower.7ex\hbox{E}\kern-.125emX}}
\begin{document}
\title{Linear Matrix Inequality Design of \textcolor{black}{Exponentially Stabilizing} Observer-Based State Feedback \textcolor{black}{Port-Hamiltonian} Controllers\break}
\author{J. Toledo, H. Ramirez, Y. Wu, and Y. Le Gorrec.
\thanks{This work has been supported by the French-German ANR-DFG INFIDHEM project ANR-16-CE92-0028 and the the EIPHI Graduate School (contract ANR-17-EURE-0002). The second author acknowledges Chilean FONDECYT 1191544 and CONICYT BASAL FB0008 projects. The third author has received founding from Bourgogne-Franche-comt\'e Region ANER 2018Y-06145. The fourth author has received funding from the European Unions Horizon 2020 research and innovation programme under the Marie Sklodowska-Curie grant agreement No 765579. }
\thanks{J. Toledo is with the Information Processing and Systems Department of the French Aerospace Lab ONERA, 2 Avenue Edouard
Belin, F-31400, Toulouse, France (e-mail: jtoledozucco@gmail.com). }
\thanks{Y. Wu, and Y. Le Gorrec are with the FEMTO-ST Institute, Univ. Bourgogne Franche-Comt\'e, ENSMM 24 rue Savary, F-25000 Besan\c{c}on, France. }
\thanks{H. Ramirez is with the Department of Electronic Engineering, Universidad T\'ecnica Federico Santa Mar\'ia, Avenida Espa\~na 1680, Valparaiso, Chile. }
%\thanks{Y. Wu is with the FEMTO-ST Institute, Univ. Bourgogne Franche-Comt\'e, ENSMM 24 rue Savary, F-25000 Besan\c{c}on, France (e-mail: yongxin.wu@femto-st.fr).}
%\thanks{Y. Le Gorrec is with the FEMTO-ST Institute, Univ. Bourgogne Franche-Comt\'e, ENSMM 24 rue Savary, F-25000 Besan\c{c}on, France (e-mail: yann.le.gorrec@ens2m.fr).}
}

\maketitle

\begin{abstract}
\textcolor{black}{The design of an observer-based state feedback (OBSF) controller with guaranteed passivity properties for port-Hamiltonian systems (PHS) is addressed using linear matrix inequalities (LMIs). The observer gain is freely chosen and the LMIs conditions such that the state feedback is equivalent to control by interconnection with \textcolor{black}{an input strictly passive (ISP) and/or an output strictly passive (OSP) and zero state detectable (ZSD) port-Hamiltonian controller are established.} It is shown that the proposed controller exponentially stabilizes a class of \jesus{infinite-dimensional} PHS and asymptotically stabilizes a class of \jesus{finite-dimensional} \jesus{non-linear} PHS. A Timoshenko beam model and a microelectromechanical system are used to illustrate the proposed approach.}
%
%The design of an observer-based state feedback controller for port-Hamiltonian (pH) systems is addressed using linear matrix inequalities (LMIs). The controller is composed of the observer and the state feedback. By passivity, the asymptotic stability of the closed-loop system is guaranteed even if the controller is implemented on complex physical systems such as the ones defined by infinite-dimensional or nonlinear models. An infinite-dimensional Timoshenko beam model and a microelectromechanical system are used to illustrate the achievable performances using such an approach under simulations.
\end{abstract}


\section{Introduction}
\label{sec:introduction}

\textcolor{black}{Port-Hamiltonian systems (PHS)} have been introduced in \cite{maschke1992port} and have shown to be well suited for the modelling and control of multi-physical systems \cite{van2000l2,duindam2009modeling}. They have been widely studied for finite-dimensional systems in \cite{van2000l2,ortega2002interconnection,ortega2004interconnection,Prajna2002}  and generalized to infinite-dimensional systems in \cite{van2002hamiltonian,LeGorrec2005,Jacob2012}. The main feature of PHS is to describe physical systems in terms of energy and to express the energy exchanges between the different internal components of these systems and their environment through an appropriate geometric structure.  

Control of PHS using Interconnection and Damping Assignment (IDA) has been proposed in \cite{ortega2002interconnection,ortega2004interconnection} and extensively developed for linear systems in \cite{Prajna2002}, in which LMIs have been employed to solve the IDA control problem. The LMIs  proposed in \cite{Prajna2002} allow to design a static state feedback that guarantees a desired closed-loop behavior. It can be seen as an alternative to traditional approaches such as pole-placement, $LQ$-control or $H_\infty$-control. Similar LMIs conditions have been used for the \textit{dual problem} of observer design in \cite{kotyczka2015dual}. %\textcolor{MinorChange}{The same LMIs are implemented for the \textit{dual problem} of the observer design in \cite{kotyczka2015dual}.} 
Further works on observer design for linear and nonlinear PHS have been reported in \cite{venkatraman2010full,vincent2016port,yaghmaei2018structure,Biedermann2018ConferencePassivity}, in which the passive properties of PHS are used to ensure the convergence of the observer. Nevertheless, few results are reported regarding OBSF control design for PHS.

In \cite{kotyczka2015dual}, the design of an OBSF controller that allows to achieve desired \jesus{closed-loop} performances and guarantees closed-loop stability when it is applied to a finite-dimensional linear system is proposed using a dual observer-based compensator. %\textcolor{MinorChange}{In that reference, the closed-loop stability is guaranteed when applying the OBSF controller to the linear pH system}. 
In \cite{wu2018reduced}, an OBSF design is proposed such that the equivalent controller is a PHS. Similarly, in \cite{Wu2020JournalReduced} the same authors proposed a similar controller for infinite-dimensional PHS with distributed actuation. However, in the two last references, stability only is considered and the closed-loop performances can only be modified through damping injection. \textcolor{black}{Recently in \cite{Toledo2020JournalObserver}, an OBSF controller has been proposed based on the resolution of a modified algebraic Ricatti equation (ARE) such that: $(i)$ the design is based on a discretized model of an infinite-dimensional PHS (as the ones presented in \cite{van2002hamiltonian,LeGorrec2005}), and $(ii)$ the \textcolor{black}{asymptotic} closed-loop stability is guaranteed when applying the finite-dimensional OBSF controller to a class of boundary controlled PHS (BC-PHS) \cite{LeGorrec2005,Jacob2012}. The design procedure consists in assuming a given state feedback gain and obtain the observer gain by solving an ARE to guarantee that the dynamic controller is \jesus{a} PHS \cite{Toledo2020JournalObserver}.%, or assuming a given observer gain and solve an ARE to obtain the stabilazing gain such that the controller is PHS \cite{Toledo2019ConferenceObserver}.
}  

\textcolor{black}{In this work we propose an extension of the preliminary results proposed in \cite{Toledo2019ConferenceObserver}. Using an early lumping approach as in \cite{Toledo2020JournalObserver}, an OBSF design method based on LMIs which guarantees that the dynamic controller is input strictly passive (ISP) and/or output strictly passive (OSP) and zero state detectable (ZSD) is proposed. The ISP, OSP and ZSD properties of the controller allow to guarantee exponential stability of the closed-loop dynamics when applied to a class of BC-PHS \cite{LeGorrec2005,Jacob2012} and the asymptotic stability when applied to a class of \jesus{non-linear} PHS \cite{van2000l2}. In both cases the design of the controller is performed on a \jesus{finite-dimensional} linear approximation of the open-loop systems and it is also possible to specify (up to a certain degree), the performance of the closed-loop \jesus{finite-dimensional} linear system whereas the closed-loop stability of the \jesus{infinite-dimensional}/non-linear system is guaranteed.} \textcolor{black}{An important improvement with respect to \cite{Toledo2020JournalObserver} and \cite{Toledo2019ConferenceObserver} is that the proposed controller \jesus{ensures} exponential stability when applied to BC-PHS, overcoming the limitation that only asymptotic stability could be achieved. The exponential stability is possible thanks to a direct feedthrough term, while the \jesus{observer-based} feedback allows to modify and improve the closed-loop response. Moreover, the LMI based design method is based on a set of explicit design parameters. An important extension of the proposed control method is that it can be equally applied to \jesus{finite-dimensional} non-linear PHS. This would also be the case for the method proposed in \cite{Toledo2020JournalObserver} provided that the \jesus{designed} controller is OSP and ZSD. \jesus{In} this paper these conditions are formalized.} 

The paper is organized as follows. Section \ref{Sec:OBSF} presents the design procedure in terms of LMIs of the proposed OBSF controller and the stability results. Section \ref{Sec:Examples} presents two examples, an infinite-dimensional Timoshenko beam model on a 1-dimensional spatial domain and a \jesus{finite-dimensional non-linear} microelectromechanical system (MEMS). The design procedure and the closed-loop performances by means of numerical simulations are illustrated. Finally, Section \ref{Section:Conclusion} gives some final remarks and discussions on possible future work.

\section{Observer-based state feedback design}\label{Sec:OBSF}
Consider the following linear PHS
\begin{equation}\label{Sys:PH ODE}
P : \begin{cases}
\dot{x}(t) &= (J-R)Q x(t) + B u(t), \,\,\,\, x(0) = x_0 \\
y(t) & = B^\top Q{x}(t)
\end{cases}
\end{equation}
where $x(t) \in \Real ^n$ is defined for all $t \geq 0$, $x_0 \in \Real ^n$ is the unknown initial condition, $u(t) \in \Real^m$ is the input and $y(t) \in \Real ^m$ is the power conjugated output of $u(t)$, which in this work is considered to be measurable. $J = -J^\top$, $R = R^\top \geq 0$ and $Q = Q^\top >0$ all known real matrices of size $n \times n$ and $B \in \Real^{n \times m}$. We assume that $J-R$ and $B$ are full rank, i.e. $rank(J-R) = n$ and $rank(B) = m$. \textcolor{black}{The Hamiltonian function of \eqref{Sys:PH ODE} is $H(t)=\frac{1}{2}x^\top(t)Qx(t)$, and for models of physical systems corresponds to the stored energy.} For simplicity, we refer to the system \eqref{Sys:PH ODE} as the system $(A,B,C)$, with $A = (J-R)Q$ and $C = B^\top Q$, and we assume that it is controllable and observable.

%\begin{assumption}}
%The PHS \eqref{Sys:PH ODE} is exponentially stable.
%\end{assumption}
%%%%%%%%%%%%%%%%%%

Define the following Luenberger observer
\begin{equation} \label{Eq:Luenberger observer}
\begin{split}
\dot{\xh}(t) &= {A_{D}} \xh(t) + B u(t) + L(y(t)- \yh(t)) + {BD_c y(t)}, \\ 
\yh(t) &= B^\top Q \xh (t), \quad \xh(0) = \xh_0,
\end{split}
\end{equation}
for the PHS \eqref{Sys:PH ODE}, where $\xh \in \Real ^n$ is the estimation of $x$, $\xh_0$ is a known initial condition,  {$A_D = A-BD_cC$ with $0<D_c \in \Real ^{m \times m}$} is a damping matrix, and $L \in \Real ^{n \times m}$ is the observer gain to be designed. {Notice that \eqref{Eq:Luenberger observer} can be interpreted as a damped version of a Luenberger observer since the design of $L$ is not based on the \jesus{open-loop} matrix $A$ but on the closed-loop matrix $A_D$.}
%An extension with respect to \cite{Prajna2002} is that the matrix $L$ is designed} such that \eqref{Eq:Luenberger observer} converges \textcolor{black}{exponentially} to \eqref{Sys:PH ODE}. 
Then, we design the feedback gain $K$ such that the following OBSF control law
\begin{equation}\label{Eq:Control Law}
\begin{split}
& u(t) = r(t) - K \xh(t)  {-D_c y(t)}, \\
&  r(t) \in \Real ^m, \quad K \in \Real ^{m \times n},
\end{split}
\end{equation}
\textcolor{black}{is equivalent to the control by interconnection with an ISP or/and OSP and ZSD\footnote{The reader is referred to the Appendix for further details on the definitions of ISP, OSP and ZSD.} PHS controller.} 
%of Figure \ref{Fig:PassiveInterconnection} in which $\hat{P}$ is a \textcolor{black}{ISP or/and OSP and ZSD controller}.
%\vspace{-1.0cm}
%leads to a closed-loop system \eqref{Eq:Luenberger observer}-\eqref{Eq:Control Law} on a pH form with inputs $r(t)$ and $y(t)$. 
%In that way %The importance of imposing this property on the observer-based state feedback controller is that 
%one can guarantee that 
Hence, if the considered linear system is the discretzation of a linear BC-PHS \cite{LeGorrec2005,Jacob2012} or the approximation of a nonlinear PHS \cite{van2000l2}, the closed-loop stability is also guaranteed if \eqref{Eq:Luenberger observer}-\eqref{Eq:Control Law} is applied to\footnote{The reader is referred to the Appendix for further details of the considered classes of systems.}:
\begin{enumerate}[label=\roman*.]
\item the original BC-PHS,
\item the original nonlinear PHS.
\end{enumerate}
%Thus, in this case t
%\textcolor{ComplicatePhrase}{The achievable closed-loop performances will also depend on the quality of the considered approximation.}
{In the following subsection we recall the LMI design approach proposed in \cite{Prajna2002} which is instrumental in our developments.}

\subsection{Observer design by LMIs}
Define the error of the state as $\xt(t) := x(t)-\xh(t)$. The error dynamics is obtained from \eqref{Sys:PH ODE} and \eqref{Eq:Luenberger observer}:
\begin{equation} \label{Eq: Error}
\dot{\xt}(t) = ( {A_D}-LC) \xt(t), \,\,\,\, \xt(0) :=  \xt_0 :=   x_0 - \xh_0,
\end{equation}
where $\xt_0$ is \textcolor{black}{an} unknown initial condition. We recall the following proposition from \cite{Prajna2002}, which is used for the design of the matrix $L$ such that $ {A_D}-LC$ is Hurwitz.

\begin{proposition}\label{Proposition:Prajna7}
\cite{Prajna2002} Denote by $\Ban$ a full rank ${(n-m) \times n}$ matrix that annihilates $B$, i.e. $\Ban  B = 0$. Let us also denote $\Ean = \Ban  {A_D}$. There exist matrices $J_d = -J_d^\top$, $R_d = R_d^\top \geq 0$, $Q_d = Q_d > 0 $ and $F$ such that $(J_d-R_d)Q_d = \jesus{A_D}+BF$ if and only if there exists a solution $\Xm = \Xm^\top \in \Real ^{n \times n}$ to the LMIs:
\begin{equation} \label{Ineq:LMI}
\begin{split}
\Xm & > 0, \\
-[\Ean \Xm \Ban ^T+ \Ban  \Xm \Ean^T ] & \geq 0.
\end{split}
\end{equation}
Given such an $\Xm$, compute $S_d$ as follows:
\begin{equation}
S_d = \begin{pmatrix}
\Ban \\ B^T
\end{pmatrix} ^{-1} \begin{pmatrix}
\Ean \Xm \\ -B^T \Xm \Ean ^T (\Ban \Ban ^T)^{-1} \Ban
\end{pmatrix},
\end{equation}
then the following matrices 
\begin{equation}\label{Eq:JdRdQdF}
\begin{array}{ll}
J_d = \tfrac{1}{2}(S_d-S_d^T), & R_d = -\tfrac{1}{2}(S_d+S_d^T), \\
Q_d = \Xm ^{-1}, & F = (B^T B)^{-1}B^T(S_d \Xm ^{-1} - {A_D} )
\end{array}
\end{equation} 
satisfy $J_d = -J_d^\top$, $R_d = R_d^\top \geq 0$, $Q_d = Q_d > 0 $ and $(J_d-R_d)Q_d =  {A_D}+BF$.
\end{proposition}
%\begin{remark}\label{Remark:Stabilizability}
%\textcolor{PossibleToDelete}{Proposition \ref{Proposition:Prajna7} is related to the stabilizability of (\ref{Sys:PH ODE}). In fact, the LMI \eqref{Ineq:LMI} has a solution if and only if the pair $(A,B)$ is stabilizable \cite[Proposition~9]{Prajna2002}. }
%\end{remark}
\begin{remark}\label{Remark:DualProblem}
The dual problem, {for the design of the matrix $L$ for the error system \eqref{Eq: Error},} consists in following Proposition \ref{Proposition:Prajna7}, but replacing {$A_D$ by $A_D^T$}, $B$ by $C^T$ and $F$ by $-L^T$. The reader can also refer to \cite[Proposition~1]{kotyczka2015dual}.
\end{remark}
%\begin{remark}
%\textcolor{PossibleToDelete}{Similar to Remark \ref{Remark:Stabilizability}, the pair $(A,C)$ is detectable if and only if the LMI \eqref{Ineq:LMI} has a solution with $\Ean=\Ban A^T$ and $\Ban \in \Real^{(n-m) \times n }$ a left annihilator of $C^T$, i.e. $\Ban C^T = 0$.}
%\end{remark}

The performances obtained using Proposition \ref{Proposition:Prajna7} are in terms of $Q_d$ (energy matrix) and $R_d$ (dissipation matrix). As it is mentioned in \cite{Prajna2002}, the LMI \eqref{Ineq:LMI} can be slightly modified in order to keep the energy matrix in a desired interval and to have sufficient but not excessive damping. This is formalized in the following corollary.
\begin{corollary}\label{Proposition:Prajna7Designing}
Under the same statements of Proposition \ref{Proposition:Prajna7}, if the following LMIs:
\begin{equation} \label{Ineq:LMIDesign}
\begin{split}
\Lambda_2 ^{-1} -\Xm  & < 0, \\
- \Lambda_1^{-1}  +\Xm & < 0, \\
\Xi_1 +\Ean  \Xm \Ban^T + \Ban  \Xm \Ean^T   & \leq 0 , \\
-\Xi_2 -\Ean  \Xm \Ban^T - \Ban  \Xm \Ean^T   & \leq 0, \\
\end{split}
\end{equation}
have a solution $\Xm = \Xm ^\top$ for some symmetric matrices $\Lambda_1$, $\Lambda_2 \in \Real ^{n \times n}$, $\Xi_1$, $\Xi_2 \in \Real^{(n-m) \times (n-m)}$,  such that $0<\Lambda_1 <\Lambda_2$ and $0\leq \Xi_1 <\Xi_2$, then $\Lambda_1<Q_d<\Lambda_2$. Moreover, choosing
\begin{equation} \label{Eq:SdMatrix}
S_d = \begin{pmatrix}
\Ban \\ B^T
\end{pmatrix} ^{-1} \begin{pmatrix}
\Ean \Xm \\ -B^T \Xm \Ean ^T (\Ban \Ban ^T)^{-1} \Ban-\gamma B^T
\end{pmatrix},
\end{equation}
for some scalar $\gamma >0$, and the matrices $J_d$, $R_d$ and $F$ as in \eqref{Eq:JdRdQdF}, then $\jesus{A_D}+BF = (J_d-R_d)Q_d$ with $R_d>0$ \textcolor{black}{which ensures exponential stability of \eqref{Eq: Error}.}
\end{corollary}
\begin{Proof}
The proof of Corollary \ref{Proposition:Prajna7Designing} is a direct application of \cite[Proposition~7, Remark~8]{Prajna2002}. See also \cite[Proposition~1]{kotyczka2015dual}.
\end{Proof}

\begin{remark} \label{Remark: AE}
Matrices $\Lambda_1$ and $\Lambda_2$ allow to fix the lowest and highest eigenvalues of $Q_d$ respectively. Matrices $\Xi_1$ and $\Xi_2$ bound the damping term, while the scalar $\gamma > 0$ implies $R_d >0$ {\cite{Prajna2002}} and then, \textcolor{black}{since $Q_d>0$, exponential behavior is ensured}.
\end{remark}

In the following section, we consider the Luenberger observer \eqref{Eq:Luenberger observer} already designed by Corollary \ref{Proposition:Prajna7Designing} using the \textit{dual problem}, i.e. replacing $\jesus{A_D}$ by {$A_D^T$}, $B$ by $C^T$, and $L = -F^T$. Then, we design the matrix $K$ of the control law \eqref{Eq:Control Law} such that the closed-loop system is equivalent to control by interconnection between the plant \eqref{Sys:PH ODE} and the OBSF controller \eqref{Eq:Luenberger observer}-\eqref{Eq:Control Law}.

\subsection{PH observer-based state feedback controller}
\begin{figure}[H]
\begin{center}
\includegraphics[width=0.25\textwidth]{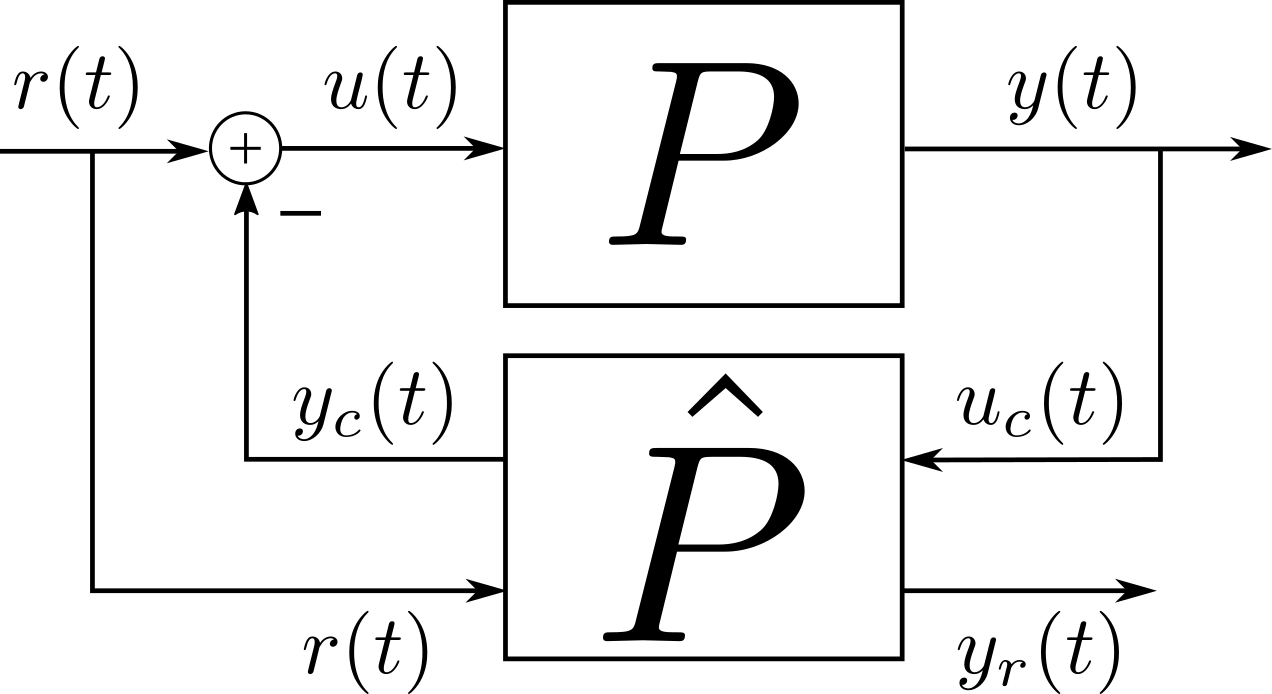}    % The printed column  
\caption{Power preserving interconnection.}  % width is 8.4 cm.
\label{Fig:PassiveInterconnection}                                 % Size the figures 
\end{center}
\end{figure}

%Consider the Luenberger observer \eqref{Eq:Luenberger observer} combined with the state feedback \eqref{Eq:Control Law}. 
The following proposition formulates the OBSF as the power preserving interconnection of \eqref{Sys:PH ODE} with a dynamic PH control system \jesus{with inputs $u_c, \, r \in \Real ^m$ and outputs $y_c, \, y_r \in \Real^m$}  as depicted in Figure \ref{Fig:PassiveInterconnection}. 
\begin{proposition}\label{Prop:Equivalent}
The OBSF \eqref{Eq:Luenberger observer}-\eqref{Eq:Control Law} is equivalent to the interconnection defined by
\begin{equation}\label{Eq:Interconnection}
u(t) = r(t)-y_c(t), \;\;\;\;\;\; u_c(t)  = y(t)
\end{equation}
between \eqref{Sys:PH ODE} and the PHS
\begin{equation}\label{Sys:Controller}
\hat{P} : 
\begin{cases}
\dot{\hat{x}}(t) &= (J_c-R_c)Q_c \hat{x}(t) + B_c u_c(t) +B r(t), \\
y_c(t) & = B_c^\top Q_c \hat{x}(t) + \textcolor{black}{D_c u_c(t)}, \\
y_r(t) & = B^\top Q_c \hat{x} (t).
\end{cases}
\end{equation}
\textcolor{black}{with $H_c=\frac{1}{2}\hat{x}^\top Q_c \hat{x}$}, if $K = B_c^T Q_c$, $B_c = L$ and the following matching equation
\begin{equation}\label{Eq:MatchCondition}
\textcolor{black}{ {A_D}}-LC-BK = (J_c-R_c)Q_c
\end{equation}
is satisfied for some matrices $J_c=-J_c^\top$, $R_c=R_c^\top\geqslant 0$, $Q_c=Q_c^\top>0$ of \textcolor{black}{appropriate dimensions}. 
\end{proposition}
\begin{Proof}
Consider in a first instance only the feedthrough term {$D_cu_c(t)$} in \eqref{Sys:Controller}. The interconnection \eqref{Eq:Interconnection}, {with $r(t)=0$,} produces the following closed-loop PHS
\begin{equation*}
\dot{x}=(A-BD_cB^\top Q)x=(J-(R+BD_cB^\top)Qx
\end{equation*}
where $(R+BD_cB^\top) = (R+BD_cB^\top)^\top \geq 0$, since $D_c = D_c^\top > 0$. Considering the complete output of \eqref{Sys:Controller}, since $K = B_c^T Q_c$ and $B_c = L$, \eqref{Eq:Luenberger observer}-\eqref{Eq:Control Law} is equivalent to \eqref{Eq:Interconnection}-\eqref{Sys:Controller} if \eqref{Eq:MatchCondition} is satisfied. 
\end{Proof}

%%%%%%%%%%%%%%%%%
\textcolor{black}{The open-loop system \eqref{Sys:PH ODE} is not necessarily exponentially stable since $R \geq 0$. However since the system is assumed controllable, and since the output is conjugated to the input, it becomes exponentially stable with the negative output feedback $u=-D_c y = -D_c B^\top Qx$, where $D_c=D_c^\top > 0$. The closed-loop system is then $\dot{x} = (J-(R+B D_c B^\top))Q x$ where $(J-(R+B D_c B^\top))Q$ is Hurwitz \cite{van2000l2}. Hence, the direct feedthrough term in \eqref{Sys:Controller} guarantees the exponential decay of the solutions of \eqref{Sys:PH ODE}. This result also holds in the case of BC-PHSs defined on 1D spatial domains \cite{LeGorrec2005,Jacob2012} provided that the boudary controller is ISP \cite{Ramirez_TAC_2014}. In the case of lumped non-linear system the use of a negative output feedback renders the system asymptotically stable provided that La Salle's invariance principle is satisfied at the equilibrium \cite{van2000l2}.}
%%%%%%%%%%%%%%%%%
\textcolor{black}{
\begin{corollary}\label{Prop:LMI}
\jesus{Consider} {$(A,B,C)$}, {$0<D_c \in \Real^{m \times m} $} and a matrix $L\in \Real^{n \times m}$ such that {$A_L:=A_D-LC$} is Hurwitz \jesus{with $A_D = A-BD_c C$}. \jesus{The} OBSF controller \eqref{Eq:Luenberger observer}-\eqref{Eq:Control Law} is equivalent to the PHS (\ref{Sys:Controller}) by computing a solution $\Xm=\Xm^\top$ to the LMIs
\begin{equation}\label{Ineq:LMI1}
\begin{split}
2 \Gamma_1 - B L^\top-L B^\top +  A_L \Xm +\Xm A_L^\top  & \leq 0, \\
-2 \Gamma_2  + B L^\top+L B^\top -  A_L \Xm - \Xm A_L^\top  & \leq 0, \\
-\Delta_1 ^{-1} + \Xm  & \leq 0, \\
\Delta_2 ^{-1} - \Xm  & \leq 0, 
\end{split}
\end{equation}
for some $n \times n$ symmetric matrices $\Gamma_1$, $\Gamma_2$, $\Delta_1$ and $\Delta_2$ such that $0 \leq \Gamma_1 < \Gamma_2$ and $0 < \Delta_1 < \Delta_2$. Then, with $S_c =  A_L \Xm - B L^\top$, one can obtain $J_c = \frac{1}{2}(S_c-S_c^\top)$, $R_c = -\frac{1}{2}(S_c+S_c^\top)$, $Q_c = \Xm ^{-1}$, $B_c = L$ and $K = B_c^\top Q_c$. Furthermore, the following results \jesus{hold:}
\begin{enumerate}[label=(\roman*)]
\item Matrices $R_c$ and $Q_c$  satisfy 
\begin{enumerate}
\item ${\Gamma_1} \leq R_c \leq {\Gamma_2} $;
\item $\Delta_1 \leq Q_c \leq \Delta_2 $;
\end{enumerate}
\item If  $\Gamma_1 >0$, then the observer estimation error converges exponentially to zero and (\ref{Sys:Controller}) is ISP, OSP and ZSD with respect to the input/output pair $u_c/y_c$. 
\end{enumerate}
\end{corollary}
\begin{Proof}
The matching equation \eqref{Eq:MatchCondition} is satisfied by the solution of  \eqref{Ineq:LMI1} and $J_c = -J_c ^\top$, $R_c = R_c ^\top$ and $Q_c = Q_c ^\top$. To conclude that the proposed solution leads to a PHS, it must be verified that $R_c \geq 0$ and $Q_c >0$. From (\ref{Ineq:LMI1}),
\begin{equation*}
\begin{split}
2 \Gamma_1  \leq  B L^\top + L B^\top -  A_L \Xm - \Xm A_L^\top  & \leq 2 \Gamma_2,  \\
{\Delta_2}^{-1} \leq \Xm & \leq {\Delta_1}^{-1}.
\end{split}
\end{equation*}
Replacing $\Xm$, $A_L\Xm -BL^T$ by their expression with respect to $S_c$ and $Q_c$, and inverting the second inequality we obtain
\begin{equation}\label{Ineq:Proof}
\;\;\;\;\; \;\;\;\;\; \begin{array}{c}
2 \Gamma_1 \leq  -(S_c+S_c^\top)   \leq 2 \Gamma_2, \\
\Delta_1 \leq Q_c  \leq \Delta_2.
\end{array}
\end{equation}
Since $R_c=-\frac{1}{2}(S_c+S_c^\top)$ we conclude that $Q_c>0$ and $R_c \geq 0$ since $\Delta_1>0$ and $\Gamma_1 \geq 0$. Implication $(i)$ is verified replacing $R_c=-\frac{1}{2}(S_c+S_c^\top)$ in \eqref{Ineq:Proof}. Implication $(ii)$ the exponential convergence to the observer follows since $R_c \geq \Gamma_1 > 0$ and $Q_c >0$. \textcolor{black}{The ISP, OSP and ZSD properties are directly verified since $R_c>0$ and $D_c>0$ \cite[Lemma 2]{KOTTENSTETTE_2014_AUTOMATICA}}. 
It is always possible to find some positive definite matrices $\Gamma_1$, $\Gamma_2$, $\Delta_1$, and $\Delta_2$ such that the LMIs in \eqref{Ineq:LMI1} are satisfied. Indeed, since $A_L$ is assumed to be Hurwitz, for any matrix $0<Q_L \in \Real^{n \times n}$, there exists a unique $\Xm>0$ such that the following Lyapunov equation $A_{L} \Xm + \Xm A_{L}^\top +Q_L = 0$ is satisfied. Then, given some $Q_L>0$, the LMIs in \eqref{Ineq:LMI1} become $2\Gamma_1 \leq BL^\top+LB^\top + Q_L \leq 2\Gamma_2$ and $\Delta_2^{-1} \leq \Xm \leq  \Delta_1^{-1}$. By choosing $Q_L = -BL^\top-LB^\top+\alpha I_n$, with $\alpha>0$ a scalar that guarantees $Q_L >0$, the LMIs in \eqref{Ineq:LMI1} become $2\Gamma_1 \leq \alpha I_n \leq 2\Gamma_2$ and $\Delta_2^{-1} \leq \Xm \leq  \Delta_1^{-1}$ for some $\alpha>0$ and with $\Xm > 0$.  For these two last inequalities, we can always find some positive definite upper and lower bounds. A simple choice for the matrices $\Gamma_1$, $\Gamma_2$, $\Delta_1$ and $\Delta_2$ is the identity modulated by a constant.
\end{Proof}}

%\begin{proposition}\label{proposition_asymptotic_stability}
%Let (\ref{Sys:PH ODE}) be the finite-dimensional or linear approximation of
%\begin{enumerate}[label=(\roman*)]
%\item a linear one-dimensional boundary controlled pH system (BC-PHS) as defined by \eqref{Eq:PDE}-\eqref{Eq:Output}, \textit{or} \label{proposition_asymptotic_stability_1},
%\item an output strictly passive (OSP) and zero-state detectable (ZTD) finite dimensional nonlinear pH system as defined by (\ref{nonlinear_system}). \label{proposition_asymptotic_stability_2}
%\end{enumerate}
%The OBSF \eqref{Eq:Luenberger observer}-\eqref{Eq:Control Law} with gains $K$ and $L$ designed using \textcolor{red}{Proposition \ref{Prop:LMI}} asymptotically stabilizes \ref{proposition_asymptotic_stability_1} or  \ref{proposition_asymptotic_stability_2}, if $\Gamma_1 >0$. Moreover, if the storage function $H(x)$ related to \ref{proposition_asymptotic_stability_2} has a global minimum in $x = 0$ and is proper, then $x= 0$ is a globally asymptotically stable equilibrium.
%\end{proposition}

\textcolor{black}{
The main result of the paper is given in the following theorem.
\begin{theo}\label{Theo}
Let (\ref{Sys:PH ODE}) be the finite-dimensional linear approximation of
\begin{enumerate}[label=(\roman*)]
\item a linear one-dimensional BC-PHS as defined by \eqref{Eq:PDE}-\eqref{Eq:Output}, \textit{or} \label{proposition_asymptotic_stability_1},
\item an OSP and ZSD \jesus{finite-dimensional non-linear} PHS as defined by (\ref{nonlinear_system}). \label{proposition_asymptotic_stability_2}
\end{enumerate}
The OBSF \eqref{Eq:Luenberger observer}-\eqref{Eq:Control Law} with gains $K$ and $L$ designed using Corollary \ref{Prop:LMI} exponentially stabilizes \ref{proposition_asymptotic_stability_1} or asymptotically stabilizes \ref{proposition_asymptotic_stability_2} if $\Gamma_1 >0$.%Moreover, if the storage function $H(x)$ related to \ref{proposition_asymptotic_stability_2} has a global minimum in $x = 0$ and is proper, then $x=0$ is a globally asymptotically stable equilibrium.
\end{theo}}

\begin{Proof}
By Proposition \ref{Prop:Equivalent}, the OBSF \eqref{Eq:Luenberger observer}-\eqref{Eq:Control Law}
is equivalent to \eqref{Sys:Controller}-\eqref{Eq:Interconnection}. By Corollary \ref{Prop:LMI}, \eqref{Sys:Controller} is ISP, OSP and ZSD if $\Gamma_1>0$. Hence the proof of $(i)$ follows by direct application of \jesus{\cite[Theorem IV.2]{Ramirez_TAC_2014}, which states that the power-preserving interconnection of a BC-PHS defined on a 1D spatial domain and an ISP finite-dimensional system is exponentially stable,} and the proof of $(ii)$ follows by direct application of \jesus{\cite[Proposition 4.3.1]{van2000l2}, which states that the power preserving interconnection between an OSP and a ZSD systems is asymptotically stable.}
\end{Proof}

\begin{remark}
The ISP, OSP and ZSD properties of the controller are fundamental for guaranteeing the closed-loop exponential stability when dealing with the control of infinite-dimensional linear systems or the asymptotic stability when dealing with finite-dimensional nonlinear systems. The PHS allow to verify these properties by matrix conditions that can be satisfied by solving the LMIs of Corollary \ref{Prop:LMI}. {The matrices $R_c$ and $Q_c$ define, respectively, the parameters of the dissipation and the energy function of the controller, which in turn define the time constants of the controller. Hence the tuning matrices $\Gamma$ and $\Delta$ can be interpreted as bounds for the controller's time constants.} This shall be illustrated in the examples in the next section.
\end{remark}

\vspace{-0.4cm}
\section{Examples}\label{Sec:Examples}
In this section we illustrate the design approach on an infinite-dimensional Timoshenko flexible beam model and on a nonlinear model of a microelectromechanical optical switch.

%\vspace{-0.5cm}
\vspace{-0.3cm}
\subsection{Boundary control of a flexible beam}
The Timoshenko beam model describes the behavior of a thick beam in a one-dimensional spatial domain. It admits the BC-PHS formulation \eqref{Eq:PDE}-\eqref{Eq:Output} with matrices
\[
\small
\begin{split}
P_1 & = 
\begin{bmatrix} 
0 & 1 & 0& 0 \\ 
1 & 0 & 0& 0 \\ 
0 & 0 & 0& 1 \\ 
0 & 0 & 1& 0  \end{bmatrix}, \qquad
P_0= 
\begin{bmatrix}
0 & 0 & 0& -1 \\ 
0 & 0 & 0& 0 \\ 
0 & 0 & 0& 0 \\ 
1 & 0 & 0& 0  
\end{bmatrix}, \\
\Lo (\ze) & =
\begin{bmatrix}
T(\ze) & 0 & 0& 0 \\
0 & {\rho(\ze)^{-1}} & 0& 0 \\
0 & 0 & EI(\ze)& 0 \\
0 & 0 & 0& {I_\rho (\ze)^{-1}}  
\end{bmatrix},
\end{split}
\]
and state variables $z = (z_1, z_2, z_3, z_4)^\top$, defined as the shear displacement $z_1(\ze,t)= w_\ze(\ze,t) - \phi(\ze,t)$, the transverse momentum distribution $z_2(\ze,t)= \rho(\ze)w_t(\ze,t)$, the angular displacement $z_3(\ze,t) = \phi_\ze(\ze,t)$ and the angular momentum distribution $z_4(\ze,t)= I_\rho (\ze)\phi_t(\ze,t)$. $w(\ze,t)$ and $\phi(\ze,t)$ are respectively the transverse displacement of the beam and the rotation angle of neutral fiber of the beam. We use lower indexes $\ze$ and $t$ to refer to the partial derivative with respect to that index. $T(\ze)$ is the shear modulus, $\rho(\ze)$ is the mass per unit length, $EI(\ze)$ is the Young's modulus of elasticity $E$ multiplied by the moment of inertia of a cross section $I$, and $I_\rho(\ze)$ is the rotational momentum of inertia of a cross section. Note that, $T(\ze)z_1(\ze,t)$ is the shear force, ${\rho(\ze)^{-1}} z_2(\ze,t)$ the longitudinal velocity, $EI(\ze) z_3(\ze,t)$ the torque and $ {I_\rho (\ze)^{-1}} z_4(\ze,t)$ the angular velocity. 

We consider the beam clamped at the left side, i.e., ${\rho(a)^{-1}} z_2(a,t) = 0$ and ${I_\rho (a)^{-1}} z_4(a,t) =0 $ and we define the following inputs and outputs
\begin{equation*}\label{Eq:InputsOutputs}
u(t)=\begin{pmatrix}
%{\rho(a)^{-1}} z_2(a,t) \\
%{I_\rho (a)^{-1}} z_4(a,t) \\
T(b) z_1(b,t) \\
EI(b) z_3(b,t) \\
\end{pmatrix}, \,\,y(t)=\begin{pmatrix}
%-T(a) z_1(a,t) \\
%-EI(a) z_3(a,t) \\
{\rho (b)^{-1}} z_2(b,t) \\
{I_\rho(b)^{-1}} z_4(b,t) \\
\end{pmatrix}.
\end{equation*}
The energy balance is given by $\dot{H}(t) = u(t)^\top y(t)$. In this case, we have force and torque actuators at the right side of the beam and collocated transverse and angular velocity sensors. The reader is \jesus{referred} to \citep{macchelli2004modeling} for more details on the model, to \citep{LeGorrec2005,Jacob2012} for the well-posedness of this class of systems and to \cite{Villegas2009,Ramirez_TAC_2014} for \jesus{exponential} stability analysis. For simplicity, we use the following parameters of the model $T=1$ Pa, $\rho  =1$ kg.m$^{-1}$, $EI=1$ Pa.m$^4$, $I_\rho =1$ Kg.m$^2$, $a=0$ m, and $b=1$ m.
%shown in Table \ref{Table:PlantParametersTB}.
%\begin{table}[H]
%\centering
%\caption{Plant Parameters.}
%\begin{tabular}[t]{lcc}
%\toprule
%& Value & Measurement unit\\
%\midrule
%$T$ 			&		$1$	&		Pa	\\
%$\rho$ 		&		$1$	&		kg.m$^{-1}$	\\
%$EI$ 			&		$1$	&		Pa.m$^4$	\\
%$I_\rho $ 	&		$1$	&		Kg.m$^2$	\\
%$a$ 			&		$0$	&		m	\\
%$b$ 			&		$1$	&		m	\\
%\bottomrule
%\end{tabular}
%\label{Table:PlantParametersTB}
%\end{table}
%\subsubsection{Finite-dimensional discretization}

A finite-dimensional approximation of the system using the finite-differences discretization scheme on staggered grids proposed in \cite{Trenchant2018} is
%This is a structure preserving spatial approximation method which preserves the pH structure of the system.
%The matrices of the finite-dimensional approximation on the form \eqref{Sys:PH ODE} are
\begin{equation*}
\scriptsize
J = \left[ \begin{array}{cccc}
0 & D & 0 & -F \\
-D^\top & 0 & 0 & 0 \\
0 & 0 & 0 & D \\
F^\top & 0 & -D^\top & 0 \\
\end{array}\right], \,\,\,\, R = 0, \\
\end{equation*}
\begin{equation*}
\scriptsize
%\begin{split}
Q = \left[ \begin{array}{cccc}
hQ_1 & 0 & 0 & 0 \\
0 & hQ_2 & 0 & 0 \\
0 & 0 & hQ_3 & 0 \\
0 & 0 & 0 & hQ_4 \\
\end{array}\right], \;
B = \left[ \begin{array}{cccc}
b_{11} & b_{12} & 0 & 0 \\
0 & 0 & b_{23} & 0 \\
0 & b_{32} & 0 & 0 \\
0 & 0 & b_{43} & b_{44} \\
\end{array}\right]
%\end{split}
\end{equation*}
where
\begin{equation*}
\scriptsize
%\begin{split}
D = \dfrac{1}{h^2}\left[ \begin{array}{cccc}
1 & 0 & \dots & 0 \\
-1 & 1 & \ddots & 0  \\
\vdots & \ddots & \ddots & \ddots \\
0 & 0 & \dots & 1 \\
\end{array}\right], \;
F =\dfrac{1}{2h}\left[ \begin{array}{cccc}
1 & 0 & \dots & 0 \\
1 & 1 & \ddots & 0  \\
\vdots & \ddots & \ddots & \ddots \\
0 & 0 & \dots & 1 \\
\end{array}\right],
%\end{split}
\end{equation*}
$Q_i,\; i \in \left\{1,\cdots,4\right\}$ are diagonal matrices containing the evaluation of  $T(\ze)$, ${\rho(\ze)^{-1}}$, $EI(\ze)$ and ${I_\rho(\ze)^{-1}}$ respectively, at the specific discretization points and
\begin{equation*}
\scriptsize
b_{11}=\tfrac{1}{h}\left[ \begin{array}{cccc}
-1 & 0 & \cdots  & 0
\end{array}\right]^\top, \,\,
b_{12}=\tfrac{1}{2}\left[ \begin{array}{cccc}
-1 & 0 & \cdots  & 0 
\end{array}\right]^\top, 
\end{equation*}
\begin{equation*}
\scriptsize
b_{23}=\tfrac{1}{h}\left[ \begin{array}{cccc}
0 & 0 & \cdots  & 1
\end{array}\right]^\top, \,\,
 b_{43}=\tfrac{1}{2}\left[ \begin{array}{cccc}
0 & 0 & \cdots  &1 
\end{array}\right]^\top,
\end{equation*}
%\begin{equation*}
%\scriptsize
%b_{11}=\dfrac{1}{h}\left[ \begin{array}{c}
%-1 \\
%0 \\
%\vdots  \\
%0 \\
%\end{array}\right], \,\,\,\,
%b_{12}=\dfrac{1}{2}\left[ \begin{array}{c}
%-1 \\
%0 \\
%\vdots  \\
%0 \\
%\end{array}\right], \,\,\,\, b_{32} = b_{11},
%\end{equation*}
%\begin{equation*}
%\scriptsize
%b_{23}=\dfrac{1}{h}\left[ \begin{array}{c}
%0 \\
%0 \\
%\vdots  \\
%1 \\
%\end{array}\right], \,\,\,\, b_{43}=\dfrac{1}{2}\left[ \begin{array}{c}
%0 \\
%0 \\
%\vdots  \\
%1 \\
%\end{array}\right], \,\,\,\, b_{44}=b_{23}.
%\end{equation*}
$b_{32} = b_{11}$ and $b_{44}=b_{23}$. The state variables of the approximated model are $x(t)=(x_1^d, x_2^d, x_3^d, x_4^d)^\top$, where $x_i^d(t)\in \mathbb{R}^{n_d}$, $i \in \left\{ 1,\cdots,4\right\}$ and the $i-th$ component of $x_1^d$, $x_2^d$, $x_3^d$ and $x_4^d$ correspond to the approximation of $z_1((i-0.5)h,t)$, $z_2(ih,t)$, $z_3((i-0.5)h,t)$ and $z_4(ih,t)$ respectively, with $h= 2\tfrac{b-a}{2*n_d+1}$, $b-a$ being the length of the beam and $n_d$ the number of elements. In this example, we choose $n_d = 10$ and hence the complete state is composed of $40$ elements.

{The numerical simulations are performed in a time interval $t = [0, 15s] $ with step time $\delta_t = 0.1 ms$ using a mid point temporal discretization \cite{Trenchant2018}. We use $100$ elements per state variable to describe the infinite-dimensional system ($400$ state variables). The observer only uses $10$ elements per state variable ($40$ state variables). The initialization is such that the beam is in equilibrium position with a force of $0.01 \;N$ applied at the end tip, i.e. $z_1(\ze,0) = 0.01$, $z_2(\ze,0) = 0$, $z_3(\ze,0) = -0.01 (\ze -1)$ and $z_4(\ze,0) = 0$. The initial condition of the observer is $\hat{x}(0) = 0$. The end-tip deformations are reconstructed from the state variables $z(\zeta,t)$ and $\hat{z}(\zeta,t)$, taking into account that the beam is clamped at the left side.}

% The reader is refereed to \cite{Trenchant2018} for further details about this discretization method.
 
{First, we design the direct feedthrough term. For simplicity, we choose $D_c = \left( \begin{smallmatrix} d_{c1} &0 \\ 0 & d_{c2} \end{smallmatrix}\right)$ with $d_{c1} = d_{c2} = 0.1$. Now, we design the matrix $L$ such that $A_D-LC$ is a Hurwitz matrix. For this purpose the LMI approach proposed in Corollary \ref{Proposition:Prajna7Designing} and a Linear Quadratic state Estimator (LQE) are used. The parameters used on Corollary \ref{Proposition:Prajna7Designing} are shown in Table \ref{Table:ObserverDesignTB}. For the LQE the following cost function
\[
J = \int_0^t\left( \xt^T \bar{Q} \xt + y^T\bar{R} y + 2\xt^T \bar{N} y\right)dt
\] 
is minimized with the design parameters shown in Table \ref{Table:ObserverDesignTB}. The matrices $C_w$, $C_\phi$, $C_1$, and $C_2$ are the output matrices related to the end-tip position, angle, transverse velocity and angular velocity, respectively. This is, $w(b,t) \approx C_w x(t)$, $\phi(b,t) \approx C_\phi x(t)$, $\rho(b)^{-1}z_2(b,t) \approx C_1 x(t)$, and $I_\rho(b)^{-1}z_4(b,t)\approx C_2 x(t)$.  
\begin{remark}
Two differences \jesus{are observed} between the design method provided in this paper and the one in \cite{Toledo2020JournalObserver}: $(i)$ The design is based on $A-BD_cC$ and not directly on $A$, \jesus{allowing to guarantee exponential instead of asymptotic stability, and} $(ii)$ \jesus{in contrast to the approach adopted in \cite{Toledo2020JournalObserver}}, the design of the matrix $L$ is free. 
\end{remark}}
\begin{table}[h]
\centering
\vspace{-0.25cm}
\caption{Observer design parameters}
\begin{tabular}[t]{cc}
LMI approach \\
\toprule
Matrix & Value \\
\midrule
$\Lambda_1$ 	&	 $0.1 I_{40}$		\\
$\Lambda_2$ 	&	 $5000 I_{40}$	\\
$\Xi_1$ 	&	 $1 I_{38}$		\\
$\Xi_2$ 	&	 $1000I_{38}$		 \\
$\gamma$ 	&	 $10$	\\
\bottomrule \\
LQE approach  \\
\toprule
Matrix & Value \\
\midrule
$\bar{Q}$ 	&	 $\gamma_w C_{w}^TC_{w}+ \gamma_\phi C_{\phi}^TC_{\phi} + \gamma_1 C_{1}^TC_{1}+ \gamma_2 C_{2}^TC_{2}$		\\
$\bar{R}$ 	&	 $ I_{2}$	\\
$\bar{N}$ 	&	 $0$		\\
& $\gamma _w = \gamma _\phi = 1000, \quad \gamma_1 = \gamma_2 = 100.$\\
\bottomrule
\end{tabular}
\label{Table:ObserverDesignTB}
\end{table}
%\begin{table}[h]
%\centering
%\caption{OBSF design parameters}
%\begin{tabular}[t]{lc}
%\toprule
%Matrix & Value \\
%\midrule
%$\Lambda_1$ 	&	 $0.1 I_{40}$		\\
%$\Lambda_2$ 	&	 $5000 I_{40}$	\\
%$\Xi_1$ 	&	 $1 I_{38}$		\\
%$\Xi_2$ 	&	 $1000I_{38}$		 \\
%$\gamma$ 	&	 $10$	\\
%$\Gamma_1$ 	&	 $0.07 I_{40}$ \\
%$\Gamma_2$ 	&	 $700 I_{40} $ 	\\
%$\Delta_1$ 	&	 $0.0064 I_{40}$ (Design 1) and $0.016 I_{40}$ (Design2)\\
%$\Delta_2$ 	&	 $200 I_{40}$	 \\
%\bottomrule
%\end{tabular}
%\label{Table:ObserverDesignTB}
%\end{table}

\begin{figure}[h]%[!htbp]
\begin{center}
\includegraphics[width=0.48\textwidth]{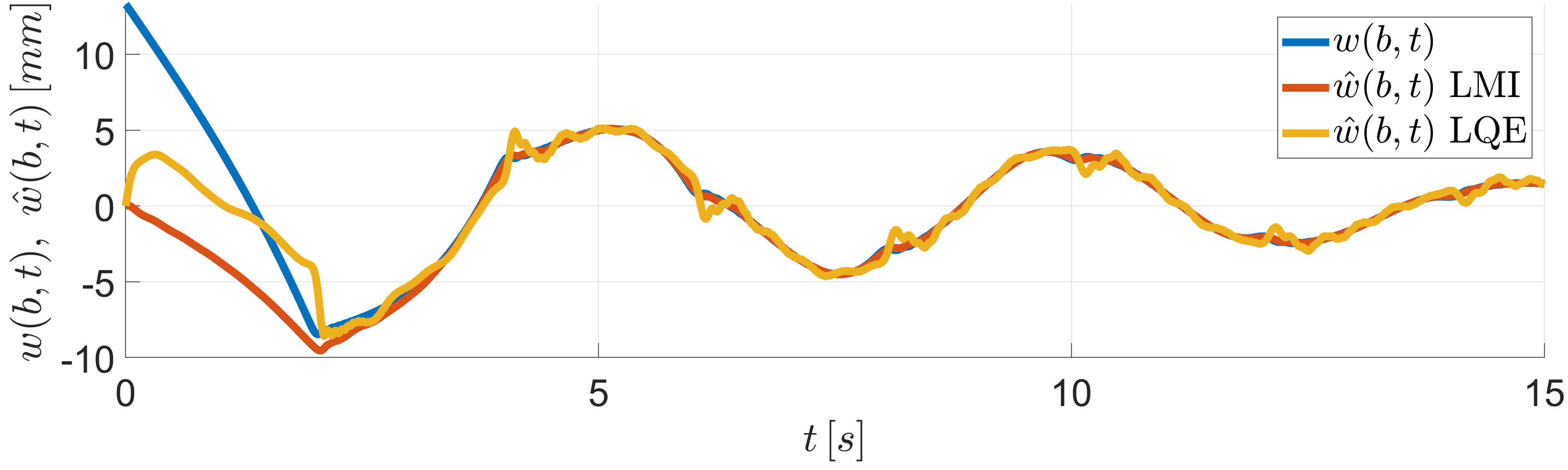}    % The printed column  
\caption{{$w(b,t)$ is the end-tip displacement using the output feedback $u(t)=-D_c y(t)$. $\hat{w}(b,t)$ is the observed end-tip position with $L$ designed using LMI and LQE methods.}}  % width is 8.4 cm.
\label{Fig:Openloop}                                 % Size the figures 
\end{center}
%\vspace{-1.0cm}
\end{figure}

Figure \ref{Fig:Openloop} shows simulations of the end-tip of the beam and \jesus{the} observed values in \jesus{closed-loop} using the feedthrough term only, i.e., $u(t) = -D_cy(t)$. Due to the freedom on the choice of $L$, we can assign completely different behaviors to the observer. In this case, \jesus{the weights $\gamma_{\omega}$ and $\gamma_{\phi}$ of the matrices $C_{\omega}$ and $C_{\phi}$, respectively, are ten times higher than the other weights (see Table \ref{Table:ObserverDesignTB}) in the LQE method. As consequence the estimation of the end-tip position is fast but sensitive to changes in the states.}

{Now, Corollary \ref{Prop:LMI} is used to design the OBSF matrix $K$. We use the design parameters shown in Table \ref{Table:FeedbackDesignTB}. For simplicity, we consider identity matrices multiplied by a positive value. Note that, for the observer designed previously with the LQE method, we add a weight value that modifies the output term ($\gamma_i C^TC$ with $i =\lbrace 3,4,5,6 \rbrace$). The last can also be applied to the LMI approach, but for simplicity we keep it as identities matrices multiplied by a coefficient.   }

\begin{table}[h]
\centering
\vspace{-0.25cm}
\caption{State feedback design parameters}
\begin{tabular}[t]{clll}
\toprule
Matrix & LMI approach & LQE approach  \\
\midrule
$\Gamma_1$ 	&	 $0.001 I_{40}$ &	 $0.001 I_{40} + \gamma_3 C^TC$	& $\gamma_3 =320 $	\\
$\Gamma_2$ 	&	 $1000 I_{40}$ &	 $1000 I_{40}+ \gamma_4 C^TC$ & $\gamma_4 = 0 $	\\
$\Delta_1$ 	&	 $0.002 I_{40}$ 	&	 $0.0002 I_{40}+ \gamma_5 C^TC$	& $\gamma_5 = 0.03 $	\\
$\Delta_2$ 	&	 $1I_{40}$ &	 $1I_{40}+ \gamma_6 C^TC$& $\gamma_6 = 0 $		 \\
\bottomrule 
\end{tabular}
\label{Table:FeedbackDesignTB}
\end{table}

\begin{figure}[h]%[!htbp]
\begin{center}
\includegraphics[width=0.48\textwidth]{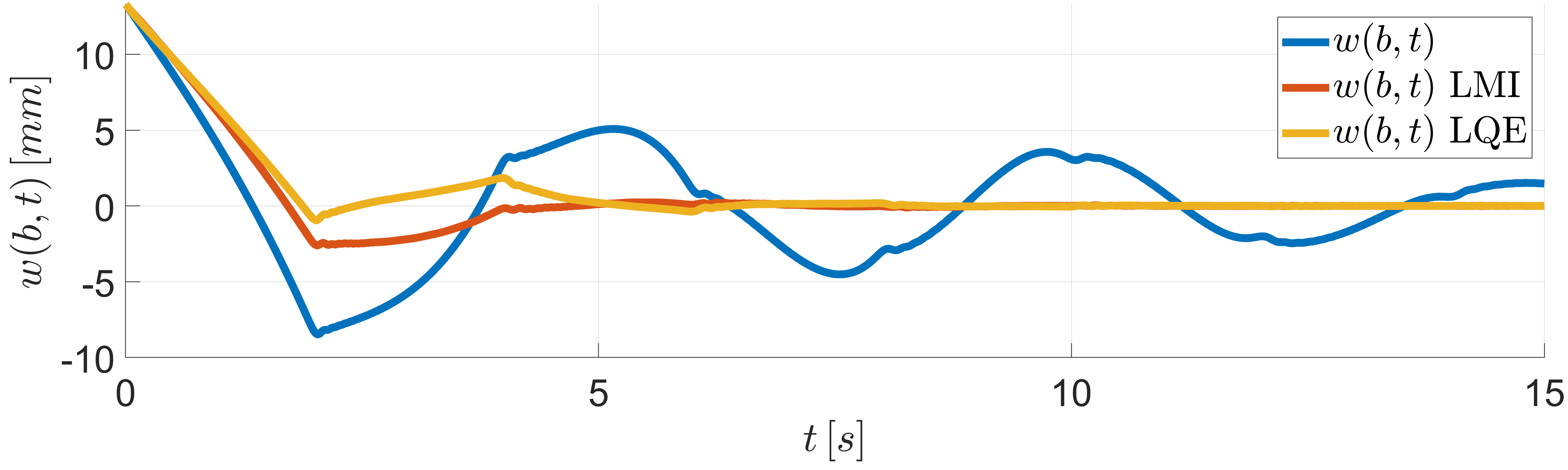}    % The printed column  
\caption{{$w(b,t)$ represents the end-tip displacement of the beam. The control law uses the feedthrough term only, the OBSF designed with the LMI approach, and the OBSF designed with the LQE approach.}}  % width is 8.4 cm.
\label{Fig:Closedloop}                                 % Size the figures 
\end{center}
\vspace{-0.5cm}
\end{figure}

{Figure \ref{Fig:Closedloop} \jesus{shows} the end-tip responses of the beam using the feedthrough term only (blue), the \jesus{closed-loop} response using the LMI approach for $L$ and $K$ (red), and the LQE approach for $L$ with the LMI approach for $K$ (yellow). We see that even if the behavior of the observer varies from both designs, by tuning the design parameters of the state feedback properly, the closed-loop responses become very similar.}

{Figure \ref{Fig:C_Closedloop} shows the influence of the design parameter $\Delta_1$ in the temporal response of the end-tip position of the beam. {Increasing the value of $\Delta_1$ reduces the settling time until the controller becomes overdamped.}
\begin{figure}[h]%[!htbp]
\begin{center}
\includegraphics[width=0.48\textwidth]{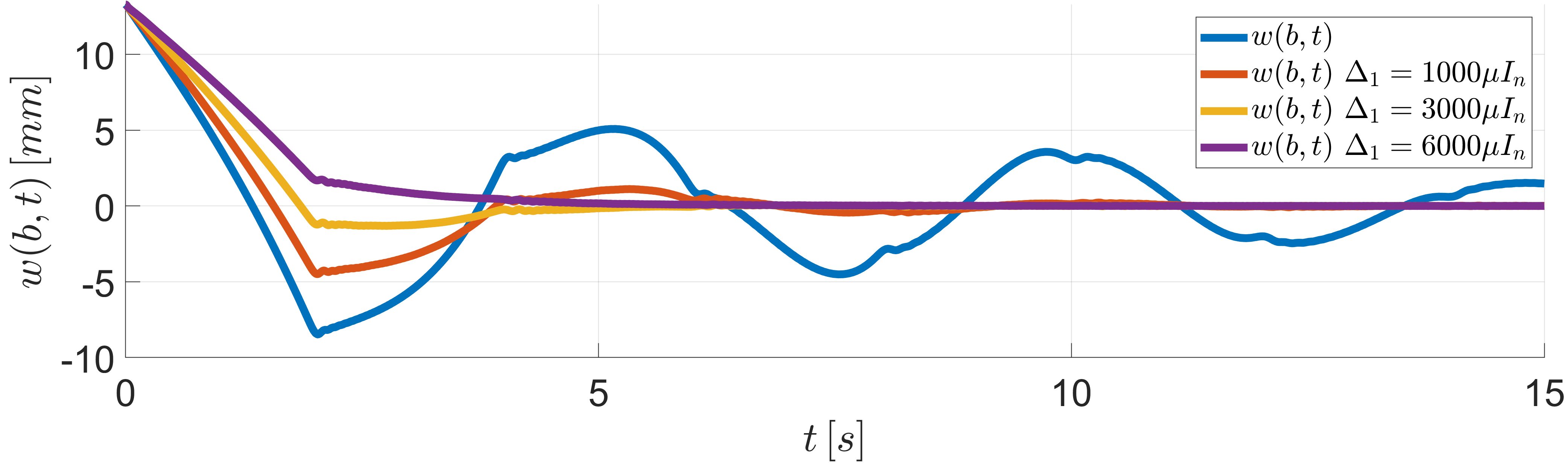}    % The printed column  
\caption{{End-tip responses for different values of the design parameter $\Delta_1$.}}  % width is 8.4 cm.
\label{Fig:C_Closedloop}                                 % Size the figures 
\end{center}
\vspace{-0.5cm}
\end{figure}
{By Theorem \ref{Theo} the reduced order OBSF controller exponentially stabilizes the BC-PHS.}
\vspace{-0.3cm}
\subsection{Microelectromechanical optical switch}

Microelectromechanical systems (MEMS) are micro robots with an electronic actuation part. Due to the miniaturization of technology, MEMS are an important tool in the micro-robotic industry. In optics for instance \cite{borovic2004control}, using tiny mirrors MEMS allow to connect two optical devices without converting continuous signals into electronic ones. A dynamical model of this system can be found in \cite{borovic2004control}  and its port-Hamiltonian representation in \cite{venkatraman2010energy}
\begin{equation}\label{Sys:MEMS}
\begin{split}
\begin{pmatrix}
\dot{q} \\ \dot{p} \\ \dot{Q} \end{pmatrix}
&= \begin{pmatrix}
0 & 1 & 0 \\ -1 & -b & 0 \\ 0 & 0 & \tfrac{-1}{r}
\end{pmatrix} \begin{pmatrix}
\tfrac{\partial H}{ \partial q} \\ \tfrac{\partial H}{ \partial p}  \\ \tfrac{\partial H}{ \partial Q} 
\end{pmatrix} + \begin{pmatrix}
0 \\ 0 \\ \tfrac{1}{r} 
\end{pmatrix} u, \\
y & = \tfrac{1}{r} \tfrac{\partial H}{ \partial Q}, \quad  C(q) = \frac{\varepsilon A_s}{q_{max}-q}, \\
 H &= \frac{p^2}{2m} + \frac{1}{2}k_1 q^2 + \frac{1}{4}k_2q^4+ \frac{Q^2}{2C(q)},
\end{split}
\end{equation}
where $q(t)$, $p(t)$ and $Q(t)$ are respectively, the position, the momentum, and the charge in the capacitor, $k_1$ and $k_2$ are the spring coefficients, $m$ is the mass of the moving part, $C(q)$ is the nonlinear capacitance which depends on the gap of the MEMS, $b >0$ and $r>0$ are the damping and resistance constant parameters, respectively, $\varepsilon$ is the dielectric constant, $A_s$ is the surface of the MEMS and $q_{max}$ is such that $q<q_{max}$. The input of the system $u(t)$ is the input voltage and $y(t)$ is the supplied current. The balance equation of the Hamiltonian is
$\dot{H}(t) = -b \left(\frac{p(t)}{m} \right)^2 - r y(t)^2 + y(t) u(t)$, which implies that the system is OSP. Under realistic operation conditions we can assume that the state space of the system is such that $Q(t)>0$ for all $t>0$, allowing to conclude that the system is ZSD. The parameters of the plant are shown in Table \ref{Table:PlantParameters}.
\begin{table}[]
%\vspace{-0.3cm}
\centering
\caption{Plant parameters and linearization point}
\vspace{-0.3cm}
\begin{tabular}[t]{lcc}
\toprule
& Value & Measurement unit\\
\midrule
$k_1$ 	&		$0.46$	&		Nm$^{-1}$	\\
$k_2$ 	&		$0.46$	&		Nm$^{-3}$	\\
$m$ 	&		$2.4 \times 10^{-8}$	&		kg	\\
$\varepsilon$ 	&		$8.854\times 10^{-12}$	&		Fm$^{-1}$	\\
$A_s$ 	&		$4 \times 10^{-4}$	&		m$^2$	\\
$q_{max}$ 	&		$  10^{-5}$	&		m	\\
$b$ 	&		$ 10^{-7}$	&		Ns	\\
$r$ 	&		$  0.5\times 10^{6}$	&		$\Omega$	\\
$\qeq$ 	&	 $0.5 \times 10 ^{-6}$		&	m	\\
$\peq$ 	&	 $0$	&	 kg m s$^{-1}$	\\
$\Qeq$ 	&	 $4.0363 \times 10^{-11}$			&	C	\\
$\ueq$ 	&	 $0.1083$		&	V \\
$\yeq$ 	&	 $2.1654 \times 10^{-8}$		& A		\\
\bottomrule
\end{tabular}
\label{Table:PlantParameters}
\end{table}
The linearization of (\ref{Sys:MEMS}) around an equilibrium point is given by 
\begin{equation}\label{Mems_linear}
\begin{split}
A &= \begin{pmatrix}
                0 &  \tfrac{1}{m} &                     0 \\
 -3 k_2 (\qeq )  ^{2} - k_1 & -\tfrac{b}{m} &  \tfrac{\Qeq}{A_s\varepsilon} \\
     \tfrac{\Qeq}{A_s \varepsilon r} &    0 & \tfrac{\qeq - q_{max}}{A_s \varepsilon r}
\end{pmatrix},  \\
B & = \begin{pmatrix}
0  \\
0  \\
 \tfrac{1}{r}
\end{pmatrix},  \qquad C = \begin{pmatrix}
 \tfrac{-\Qeq}{A_s \varepsilon r} & 0 & \tfrac{-\qeq - q_{max}}{A_s \varepsilon r}
\end{pmatrix},
\end{split}
\end{equation}
with the $\ast$ symbol representing the equilibrium point used for the linearisation (see the values in Table \ref{Table:PlantParameters}).

{For the OBSF design we choose $D_c = 10^4$. Using this value, the eigenvalues of the matrix $A_D = A-BD_cC$ are close to the eigenvalues of the matrix $A$. In a second step the matrix $L$ is designed such that $A_D-LC$ is a Hurwitz matrix. Similarly as in the previous example, we use the LMI and the LQE methods using Corollary \ref{Proposition:Prajna7Designing} by minimizing the cost function $J = \int_0^t \left(\xt^T \bar{Q} \xt + y^T\bar{R} y + 2\xt^T \bar{N} y \right)dt$, respectively. The design parameters are shown in Table \ref{Table:ObserverDesign} where the matrices $C_q$, $C_p$ and $C_Q$ are defined as $C_q = \left(\begin{smallmatrix}
1 & 0 & 0
\end{smallmatrix}\right)$, $C_p = \left(\begin{smallmatrix}
0 & 1 & 0
\end{smallmatrix}\right)$ and $C_Q = \left(\begin{smallmatrix}
0 & 0 & 1
\end{smallmatrix}\right)$.  }

%Following the design procedure of section \ref{Sec:OBSF} the linearized model \eqref{Mems_linear} is used for the synthesis of the OBSF controller. The observer is designed using Proposition \ref{Proposition:Prajna7Designing} with the parameters given in Table \ref{Table:ObserverDesign}. The eigenvalues of the matrix $A_L= A-LC$ are shown in Figure \eqref{Fig:MEMS_A}.

\begin{table}[!h]
\centering
\vspace{-0.25cm}
\caption{Observer design parameters}
\begin{tabular}[t]{cc}
LMI approach \\
\toprule
Matrix & Value \\
\midrule
$\Lambda_1$ 	&	 $1 \times 10^{-2} \times  diag([1,200,1])$		\\
$\Lambda_2$ 	&	 $1 \times 10^{10} I_3 $	\\
$\Xi_1$ 	&	 $1 \times 10^{-1}I_2$			\\
$\Xi_2$ 	&	 $1 \times 10^{4}I_2$		 \\
$\gamma$ 	&	 $30 \times 10^{4}$	\\
\bottomrule \\
LQE approach  \\
\toprule
Matrix & Value \\
\midrule
$\bar{Q}$ 	&	 $\gamma_q C_{q}^TC_{q}+ \gamma_p C_{p}^TC_{p} + \gamma_Q C_{Q}^TC_{Q}$		\\
$\bar{R}$ 	&	 $ 1$	\\
$\bar{N}$ 	&	 $0$		\\
& $\gamma _q = \gamma _p = 8 \times 10^5, \quad \gamma_1 = \gamma_Q = 10^3$\\
\bottomrule
\end{tabular}
\label{Table:ObserverDesign}
\end{table}
{For the simulation, time $t = [0, 10\,ms]$ is used with a step time $\delta t = 1 \; \mu s$. The initial conditions are set equal to $q(0)=\qeq$, $p(0)=\peq$, $Q(0)=0.9\Qeq$ for the nonlinear system, while for the observer all initial conditions are set exactly at the equilibrium point.}
{Figures \ref{Fig:Openloop_q}, \ref{Fig:Openloop_p}, and \ref{Fig:Openloop_Qc} show the temporal responses of the position, momentum and the electric charge electric of the closed-loop system using the feedthrough term only ($u = -D_cy$). In this case the LQE method has been used to speed-up the observed position and momentum in exchange of a bigger overshoot.}

\begin{figure}[!h]%[!htbp]
\begin{center}
\includegraphics[width=0.47\textwidth]{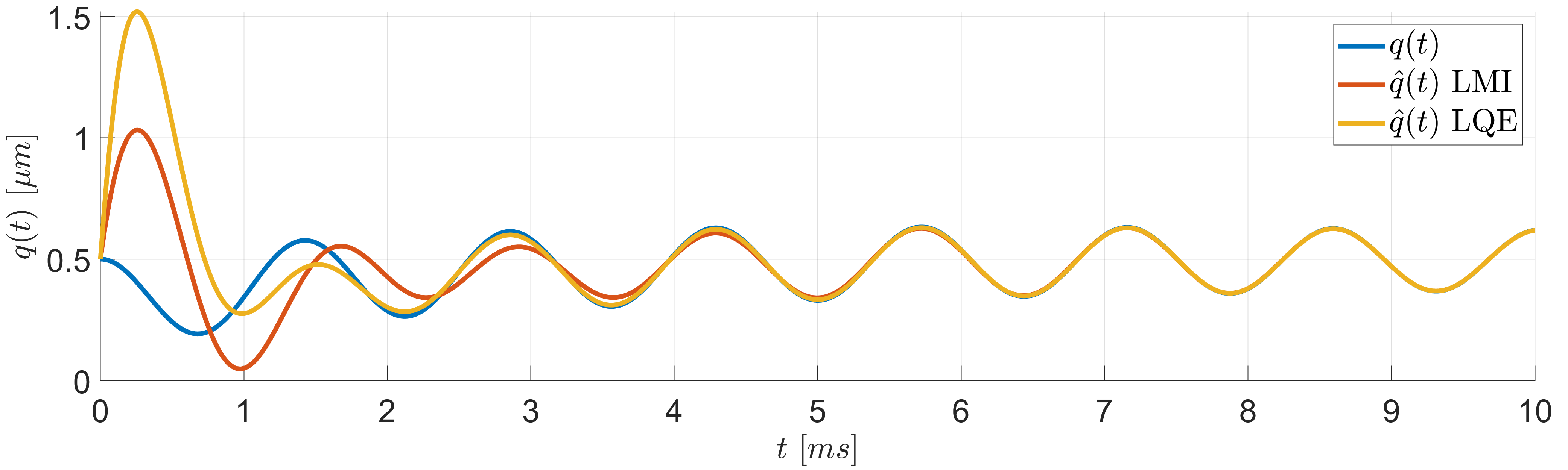}    % The printed column  
\caption{{$q(t)$ is the position and $\hat{q}(t)$ is the observed one.}}  % width is 8.4 cm.
\label{Fig:Openloop_q}                                 % Size the figures 
\end{center}
\vspace{-0.7cm}
\end{figure}
\begin{figure}[!h]%[!htbp]
\begin{center}
\includegraphics[width=0.47\textwidth]{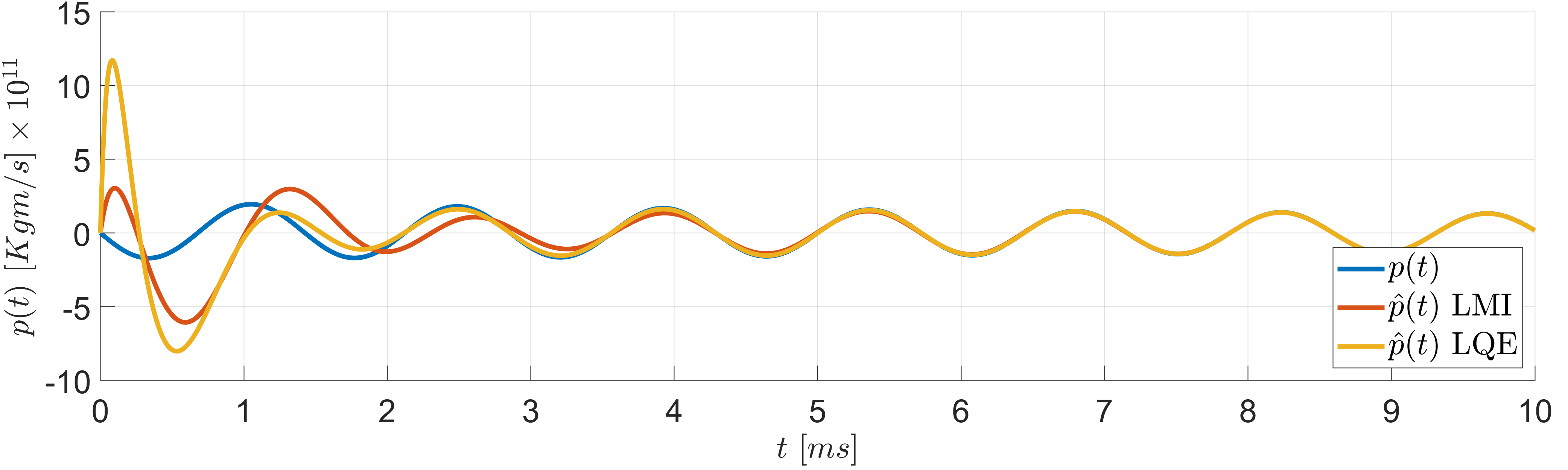}    % The printed column  
\caption{{$p(t)$ is the momentum and $\hat{p}(t)$ is the observed one.}}
\label{Fig:Openloop_p}                                 % Size the figures 
\end{center}
\vspace{-0.7cm}
\end{figure}
\begin{figure}[!h]%[!htbp]
\begin{center}
\includegraphics[width=0.47\textwidth]{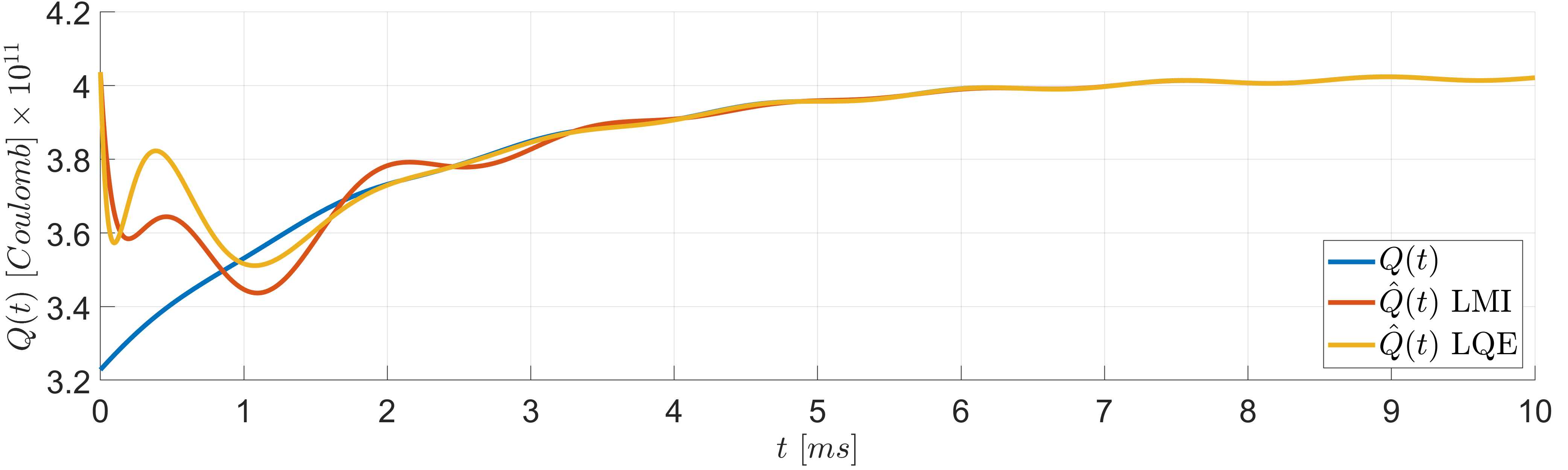}    % The printed column  
\caption{{$Q(t)$ is the electric charge and $\hat{Q}(t)$ is the observed one.}}
\label{Fig:Openloop_Qc}                                 % Size the figures 
\end{center}
\vspace{-0.3cm}
\end{figure}

{Now, for both observers a state feedback matrix $K$ \jesus{is} designed using Corollary \ref{Prop:LMI} with the design parameters shown in Table \ref{Table:FeedbackDesign}. Figure \ref{Fig:Closedloop_q} shows the dynamical responses for the position $q(t)$ (for the sake of space only the position is shown.). For both observers, one can achieve similar \jesus{closed-loop} behaviors by tuning the state feedback properly.}
\vspace{-0.3cm}
\begin{table}[!h]
\centering
\caption{State feedback design parameters}
\vspace{-0.4cm}
\begin{tabular}[t]{clll}
\toprule
Matrix & LMI approach & LQE approach  \\
\midrule
$\Gamma_1$ 	&	 $10^{-6} I_{3}$ &	 $10^{-7}I_{3} $	\\
$\Gamma_2$ 	&	 $10^{12} I_{3}$ &	 $10^{5} I_{3} $	\\
$\Delta_1$ 	&	 $1.2^{-1} I_{3}$ 	&	 $0.6^{-1}I_{3} $	\\
$\Delta_2$ 	&	 $10^{10}I_{3} $ &	 $10^{12} I_{3} $		 \\
\bottomrule 
\end{tabular}
\label{Table:FeedbackDesign}
\end{table}
\vspace{-0.4cm}
\begin{figure}[!h]%[!htbp]
\begin{center}
\includegraphics[width=0.48\textwidth]{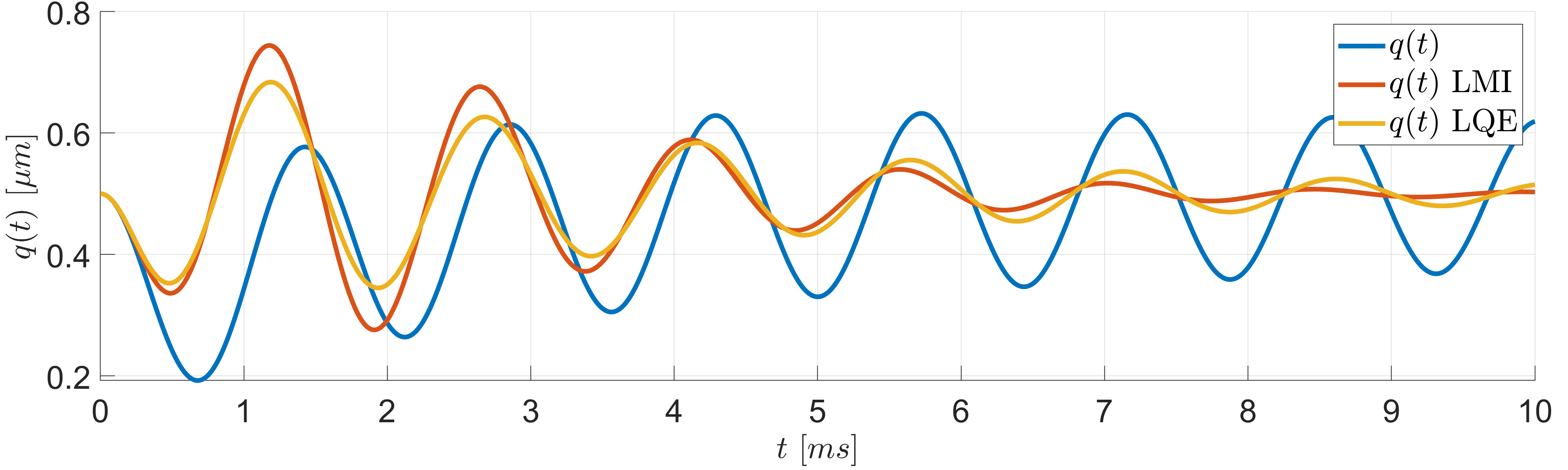}    % The printed column  
\caption{{$q(t)$ represents the position. The control law uses the feedthrough term only}, the OBSF designed with the LMI approach, and the OBSF designed with the LQE approach.}  % width is 8.4 cm.
\label{Fig:Closedloop_q}                                 % Size the figures 
\end{center}
\vspace{-0.5cm}
\end{figure}
%\begin{figure}[!h]%[!htbp]
%\begin{center}
%\includegraphics[width=0.48\textwidth]{Figures/B_ClosedLoopMEMS_p.png}    % The printed column  
%\caption{{$p(t)$ is the momentum. The open loop ($u = -D_cy$) response in blue, whereas the closed-loop ($u = -D_cy - K\hat{x}$) in red and yellow.}}\label{Fig:Closedloop_p}                                 % Size the figures 
%\end{center}
%%\vspace{-1.0cm}
%\end{figure}
%\begin{figure}[!h]%[!htbp]
%\begin{center}
%\includegraphics[width=0.48\textwidth]{Figures/B_ClosedLoopMEMS_Qc.png}    % The printed column  
%\caption{\jesus{$Q(t)$ is the electric charge. The open loop ($u = -D_cy$) response in blue, whereas the closed-loop ($u = -D_cy - K\hat{x}$) in red and yellow.}}\label{Fig:Closedloop_Qc}                                 % Size the figures 
%\end{center}
%%\vspace{-1.0cm}
%\end{figure}

%\vspace{-2cm}
{Finally, Figure \ref{Fig:Closedloop_DiffDelta1} shows the influence of $\Delta_1$ in the temporal response of the position. One can see that increasing the value of $\Delta_1$ the settling time is reduced.}
\begin{figure}[!h]%[!htbp]
\begin{center}
\includegraphics[width=0.48\textwidth]{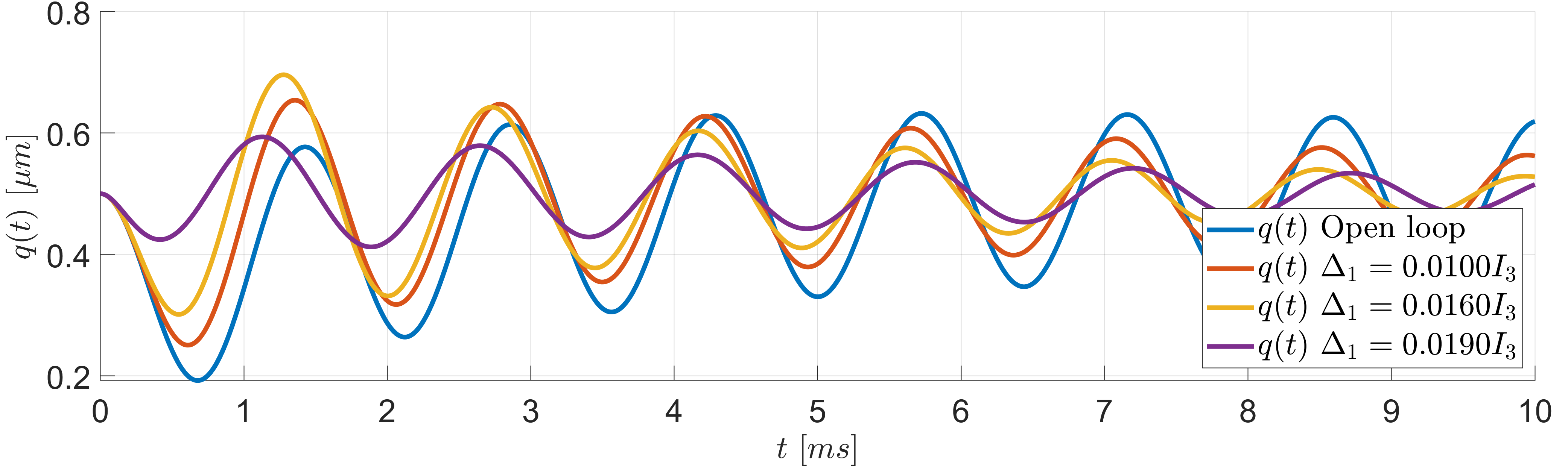}    % The printed column  
\caption{{Position responses for different values of the design parameter $\Delta_1$.}}\label{Fig:Closedloop_DiffDelta1}                                 % Size the figures 
\end{center}
%\vspace{-1.0cm}
\vspace{-0.5cm}
\end{figure}

{Theorem \ref{Theo} guarantees that the OBSF controller asymptotically stabilizes the nonlinear system \eqref{Sys:MEMS}. In general, the design procedure can be summarized as follows:
\begin{enumerate}[label=\roman*.]
\item Approximate the infinite-dimensional/non-linear model by a linear finite-dimensional model.
\item Assign the observer behavior to the finite-dimensional model. To this end, different strategies can be used, namely LMI, LQE, or pole-placement, for instance.
\item Tune the matrices $\Gamma_1$, $\Gamma_2$, $\Delta_1$, and $\Delta_2$ for the design of $K$.
\item Apply the OBSF controller to the original system.  
\end{enumerate}}

\vspace{-0.2cm}
\section{Conclusion}\label{Section:Conclusion}
An OBSF LMI based design is proposed for linear PHS. The feedback consists of a Luenberger observer and a negative feedback on the observed state. The novelty and main contribution of this paper is a constructive design method for the OBSF gains based on $(i)$ a discretized model of BC-PHS, or $(ii)$ a linearized model of a nonlinear PHS. The observer gain is designed freely and the state feedback gain is designed such that the OBSF controller is ISP and/or OSP and ZSD. The closed-loop performances can then be modified by tuning some explicit controller parameters while guaranteeing the closed-loop exponential/asymptotic stability when the OBSF is applied to the original system.
%such that we can either modify the closed-loop performances and guarantee closed-loop stability . The gain synthesis can be based either on a discretized model of a boundary controlled pH system, or on a linearized model of a nonlinear pH system. Then, we can drastically change the closed-loop performances on a range of frequencies given by the discretized model 
% to cast the feedback and the Luenberger observer as a pH system interconnected in a power preserving manner with the system to be controlled. This reinterpretation of the OBSF controller allows to use the SPR property  of the system to guarantee the closed-loop stability. 
% The second contribution of this paper is a constructive design methodology using LMIs such that one can assign desired performances on a discretized model/linearized model from an infinite-dimensional system/nonlinear system, respectively, while guaranteeing the closed-loop stability when the OBSF controller is applied to the original system.
%work is to explicitly give the conditions such that the OBSF controller is SPR, OSP, and ZSD. This conditions are satisfied by using LMIs
% use the proposed OBSF controller to asymptotically stabilize a large class of linear boundary controlled infinite dimensional pH systems and nonlinear pH systems when using a linear approximation of these system to design the controller.
An infinite-dimensional Timoshenko beam model and a finite-dimensional nonlinear model of a microelectromechanical actuator have been used to illustrate the effectiveness of the proposed approach.

\vspace{-0.25cm}\appendices
\section*{Appendix}
\subsection*{Boundary controlled pH system on 1D domain}
In this subsection the definition of Boundary Controlled Port-Hamiltonian Systems (BC-PHS) is given. The reader is refereed to \cite{LeGorrec2005,Jacob2012} for further details and definitions. A BC-PHS is a dynamical system governed by the following partial differential equation
\begin{align}
&\dfrac{\partial z}{\dt}(\zeta,t) = P_1 \dfrac{\partial}{\partial \zeta} (\mathcal{H}(\zeta) z(\zeta,t)) +P_0 \mathcal{H}(\zeta) z(\zeta,t), \label{Eq:PDE}\\ & z(\zeta,0) = z_0(\zeta), \label{Eq:IC}\\
& W_\mathcal{B} \left(\begin{smallmatrix}
\fb(t) \\
\eb(t) \\
\end{smallmatrix}\right) = u(t), \label{Eq:Input}\\
& y(t) = W_\mathcal{C} \left(\begin{smallmatrix}
\fb(t) \\
\eb(t) \\
\end{smallmatrix}\right). \label{Eq:Output}
\end{align}
where the initial condition is given by \eqref{Eq:IC}, the boundary input by \eqref{Eq:Input} and the boundary output by \eqref{Eq:Output}. Here $z(\zeta,t) \in \mathbb{R}^n$ is the state variable with initial condition $z_0(\zeta)$. $\zeta \in [a,b]$ is the 1D domain and $t \geq0$ is the time. $P_1 = P_1 ^T  \in \mathbb{R}^{n \times n}$ is a non-singular matrix, $P_0 = -P_0 ^T  \in \mathbb{R}^{n \times n}$, $\mathcal{H} (\zeta)$ is a bounded and continuously differentiable matrix-valued function satisfying for all $\zeta \in [a,b]$, $\mathcal{H} (\zeta)=\mathcal{H} ^T (\zeta)$ and $mI < \mathcal{H} (\zeta)< MI$ with $0< m < M$ both scalars independent on $\zeta$. The Hamiltonian energy function of \eqref{Eq:PDE} is given by
%\begin{equation*}\label{Eq:Hamiltonian}
$H(t) = \dfrac{1}{2} \int_a^b {z(\zeta,t)^T \mathcal{H}(\zeta) z(\zeta,t)}d\zeta.$
%\end{equation*}
$\left(\begin{smallmatrix}
\fb(t) \\
\eb(t) \\
\end{smallmatrix}\right)$ are the {\it boundary port variables} defined as
\begin{equation*}\label{B_Eq:BPV}
\begin{pmatrix}
f_\partial (t) \\
e_\partial (t)
\end{pmatrix}=\frac{1}{\sqrt{2}}\begin{pmatrix}
P_1 & -P_1 \\
I & I
\end{pmatrix}\begin{pmatrix}
\mathcal{H} (b) z(b,t) \\
\mathcal{H} (a) z(a,t) \\
\end{pmatrix}.
\end{equation*}
$W_\mathcal{B}$, $W_\mathcal{C} \in \mathbb{R}^{n \times 2n}$ are two matrices such that if $W_\Bo \Sigma W_\Bo ^T={W}_\Co \Sigma {W}_\Co ^T=0$ and ${W}_\Co \Sigma W_\Bo^T=I$, with $\Sigma = \left( \begin{smallmatrix} 0 & I \\ I & 0 \end{smallmatrix} \right)$, then $\dot{H}(t) = u(t)^T y(t)$.

\vspace{-0.8cm}
\textcolor{black}{\subsection*{ISP, OSP and ZSD (non)-linear control system}
In this subsection the definitions of input strictly passive (ISP), output strictly passive (OSP) and zero-state detectable (ZSD) (non)-linear systems are given. The reader is refereed to \citep{van2000l2} for further details and definitions. Consider a control system
\begin{equation}\label{nonlinear_system}
\dot{x}=f(x,u), \quad y=h(x,u)
\end{equation}
with $x \in \mathbb{R}^n$, $u \in \mathbb{R}^m$, $y \in \mathbb{R}^m$ and $f(\cdot)$ and $h(\cdot)$ sufficiently smooth differentiable mappings, then (\ref{nonlinear_system}) is
\begin{itemize}
\item ISP if there exists $\delta > 0$ such that it is dissipative with respect to the supply rate   $s(u,y)=u^\top y - \delta \| u \|^2$, 
\item OSP if there exists $\epsilon > 0$ such that it is dissipative with respect to the supply rate   $s(u,y)=u^\top y - \epsilon \| y \|^2$, 
\item ZSD if $u(t) = 0$, $y(t) = 0$, $\forall t \geq 0$, implies $\lim_{t\to\infty} x(t)=0$.
\end{itemize}}
%A (nonlinear) PHS is a dissipative system with storage function $H(x)$ \citep{van2000l2}. 

%\vspace{-0.5cm}
%\section*{Acknowledgment}
\vspace{-0.3cm}
\bibliographystyle{IEEEtran}
\bibliography{references_JTZ}           % and a bib file to produce the 

\end{document}